\documentclass[prl, aps, showkeys, twocolumn,
superscriptaddress, showpacs,
]{revtex4-1}
\usepackage[utf8]{inputenc}
\usepackage{verbatim}
\usepackage{amsmath}
\usepackage{amssymb}
\usepackage{graphicx} 
\usepackage{color}
\usepackage[maxfloats=100]{morefloats}

\newcommand{\E}[1]{\langle #1 \rangle}

\newcommand{\lig}{\ensuremath{\mathrm{L}}{}}
\newcommand{\rec}{\ensuremath{\mathrm{R}}}
\newcommand{\lr}{\ensuremath{\mathrm{LR}}}
\newcommand{\nmax}{\ensuremath{N}}
\newcommand{\g}{\ensuremath{\tilde{\gamma}}}

\newcommand{\A}{{\ensuremath{\cal A}}}
\newcommand{\B}{\ensuremath{{\cal B}}}

\newcommand{\peq}{\ensuremath{p_\mathrm{eq}}}
\newcommand{\Qbind}{\ensuremath{Q_\mathrm{bind}}}
\newcommand{\Qex}{\ensuremath{Q_\mathrm{exch}}}

\newcommand{\beq}{\begin{equation}}
\newcommand{\eeq}{\end{equation}}
\newcommand{\beqn}{\begin{eqnarray}}
\newcommand{\eeqn}{\end{eqnarray}}
\newcommand{\elabel}[1]{\label{eq:#1}}
\newcommand{\eref}[1]{Eq.~\ref{eq:#1}}
\newcommand{\erefs}[2]{Eqs.~\ref{eq:#1} and \ref{eq:#2}}

\newcommand{\fref}[1]{Fig.~\ref{fig:#1}}

\newcommand{\Em}{\mathcal{E}^-}
\newcommand{\Ep}{\mathcal{E}^+}
\newcommand{\Epm}{\mathcal{E}^\pm}

\newcommand{\hal}{[\alpha]}

\let\oldmarginpar\marginpar
\renewcommand\marginpar[1]{\oldmarginpar[\raggedleft\footnotesize #1]%
{\raggedright\footnotesize #1}}

\newcommand{\outline}[1]{}
\newcommand{\todo}[1]{}

\renewcommand{\todo}[1]{\marginpar{TODO: {#1}}}

\begin{document}


\author{Nils B. Becker}
\altaffiliation[Present address: ]{Bioquant, Universtit\"at Heidelberg, Im
Neuenheimer Feld 267, 69120 Heidelberg, Germany}

\author{Andrew Mugler}
\email{mugler@amolf.nl}
\altaffiliation[Present address: ]{Department of Physics, 
Emory University, Atlanta, GA 30322, USA}

\author{Pieter Rein \surname{ten Wolde}}
\email{tenwolde@amolf.nl}

\affiliation{FOM Institute AMOLF, Science Park 104, 1098 XG Amsterdam, The 
Netherlands}

\title{Prediction and Dissipation in Biochemical Sensing}

\date{\today}

\begin{abstract}  

Cells sense and predict their environment via energy-dissipating 
pathways.
However, it is unclear whether dissipation helps or harms prediction.
Here we study dissipation and prediction for a minimal sensory
module of receptors that reversibly bind ligand.
We find that the module performs short-term prediction optimally when
operating in an adiabatic regime where dissipation vanishes.
In contrast, beyond a critical forecast interval, prediction becomes
most precise in a regime of maximal dissipation, suggesting that
dissipative sensing in biological systems can serve to enhance
prediction performance.
\end{abstract}


\maketitle


The ability to sense and respond to changing environments is a defining
property of life.  An optimal response often targets future states of the
environment, either because it requires a minimum time to mount
\cite{mitchell09}, or because it inherently depends on the timing of future
events \cite{scialdone13}.  It is therefore critical not only to sense the
environment but also to predict it \cite{bialek12}.  Single cells perform
sensing by biochemical reactions, and it is natural to think that optimal
predictive power is provided by biochemical components that respond quickly. A
quick response can be achieved straightforwardly by rapid equilibration of
reactions with fast intrinsic rates, dissipating no energy.  Indeed, in a
recent study by Still \emph{et al.}, it was shown that driven systems which
dissipate minimal energy are also the most predictive \cite{still12a}.

Yet, in some cellular contexts where prediction is expected to be critical, such
as chemotaxis and circadian oscillation, components respond not quickly but
roughly on the timescale of the driving signal \cite{kaupp08, valencia12}, thus
dissipating energy.  Dissipation has been found to be beneficial \cite{skoge13}
or even essential \cite{detwiler00, qian05, mehta12a, lan12c, barato13,
govern13} for instantaneous sensing in various model systems, suggesting,
contrary to~\cite{still12a}, it may also aid prediction.

In this Letter, we address the interplay between prediction, dissipation, and
response time for a simple and ubiquitous ligand-receptor sensory module. We
find that for near-future prediction, a non-dissipative fast response is
optimal.  Surprisingly, beyond a critical prediction interval, a maximally
dissipative slower response becomes optimal.
This effect is generic and relies on non-Markovian signal autocorrelations.

\begin{figure}[tb]
\begin{center}
\includegraphics[width=\columnwidth]{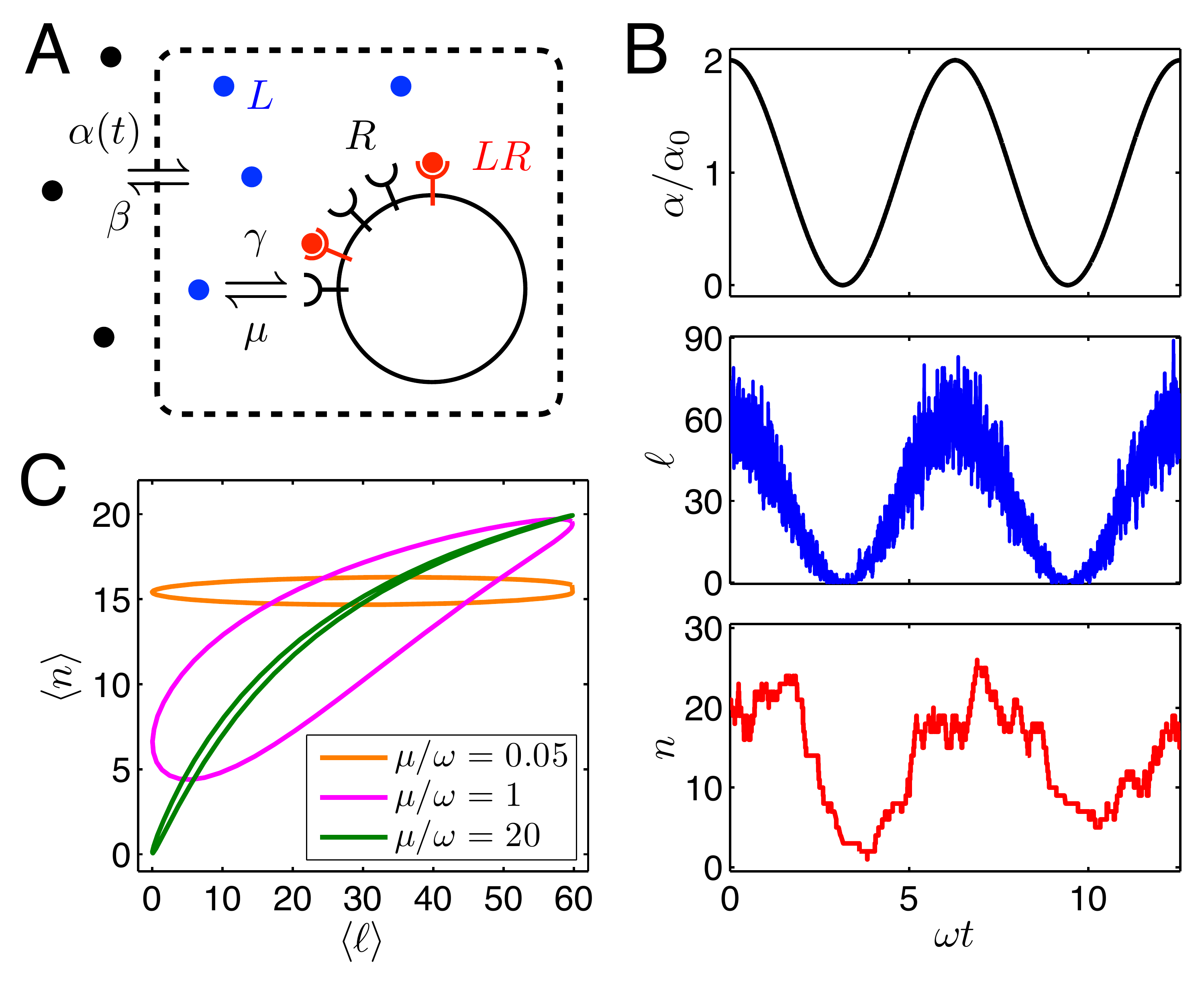} 

\caption{Biochemical sensory module. (A) Ligand, inserted dynamically and
removed, binds and unbinds receptors. (B) An oscillatory insertion rate
$\alpha(t)$ produces an oscillating ligand number $\ell(t)$, which drives the
dynamics of the bound receptor number $n(t)$. (C) The mean dynamics are
characterized by the timescale ratio $\mu/\omega$. Parameters are  $\rho = 1$,
$\lambda_0 = 30$, $\beta/\omega = 100$, $\nmax = 30$, $\gamma = \mu/\lambda_0$,
and, in B, $\mu/\omega = 1$.}

\label{fig:setup}
\end{center}
\end{figure}

\paragraph{Setup}
We consider a minimal sensory module (Fig.~\ref{fig:setup}A):
Ligand molecules \lig~are inserted into a reaction volume at rate
$\alpha(t)$ and removed with rate constant $\beta$. A pool of \nmax~receptor
molecules \rec~can bind and unbind ligands, with rate constants $\gamma$ and
$\mu$, respectively:
\begin{equation}
\label{eq:reactions}
\varnothing \underset{\beta}{\overset{\alpha(t)}{\rightleftharpoons}} \lig;\quad
\lig + \rec \underset{\mu}{\overset{\gamma}{\rightleftharpoons}}  \lr.
\end{equation}
The rate $\alpha(t)$ dynamically creates $\ell(t)$ free ligand molecules, which
drive the formation of $n(t)$ LR complexes, leaving $r(t)\equiv\nmax-n(t)$
receptors unbound.  $\ell(t)$ acts as a noisy reporter for $\alpha(t)$, which
is a proxy for the background ligand concentration to be sensed by the cell.
We take $\alpha(t)$ as the input and the response $n(t)$ as the output of the
module.

To study how dissipation and prediction depend on the dynamics of $\alpha(t)$,
we first consider the sinusoidal signal $\alpha(t) = \alpha_0[1+\rho\cos(\omega
t)]$, with $0 \le \rho \le 1$.  The response is shown in Fig.~\ref{fig:setup}B
and C: The free ligand number $\ell(t)$ oscillates around its mean
$\lambda_0\equiv \alpha_0/\beta$ stochastically and, for sufficiently fast
ligand exchange, in phase with $\alpha(t)$. The output $n (t)$ is damped and
lags behind the signal, depending on the binding speed $ \mu/\omega$, which is
the primary design parameter.

The stochastic dynamics are given by the Master equation
\begin{equation}
\label{eq:CME}
\partial_t p(\ell,n|t)
 = \bigl(\B_\ell^{\alpha(t),\beta}+ \A_{\ell,n}^{\gamma,\mu,\nmax}\bigr) 
 p(\ell,n|t).
\end{equation}
Here $\B_\ell^{\alpha,\beta} = \alpha(\Em_\ell-1)+\beta(\Ep_\ell-1)\ell$,
$\A_{\ell,n}^{\gamma,\mu,\nmax} = \gamma(\Ep_\ell\Em_n-1)\ell r
+\mu(\Em_\ell\Ep_n-1)n$, and $\Epm_xf(x) = f(x_\pm)$ define the birth-death
ligand exchange, ligand association, and step operators, respectively, and
$x_\pm\equiv x \pm1$. Appropriate boundary conditions enforce $\ell\ge0$ and
$0\le n\le \nmax$. 

\paragraph{Dissipated heat} 

To characterize the thermodynamics
\cite{nicolis77, schnakenberg76, gaspard04, hill04, schmiedl07a} of our 
module, we require all reactions to be elementary, i.e.~without implicit
dissipative steps such as ATP turnover \cite{seifert12b}.
If $\alpha(t)=\alpha$ were a constant, the stationary solution of
Eq.~\ref{eq:CME} would take the thermodynamic equilibrium form
\cite{kampen92}
$ \peq(\ell, n) \propto z^\ell_\lig z^r_\rec z^n_\lr/(\ell!r!n!) $,
where detailed balance requires $z_\lig = \alpha/\beta$ and 
$z_\lr/z_\rec = \alpha\gamma/(\beta\mu)$, and $z_\rec$ is fixed by
normalization. The Gibbs free energy $\Phi = -\log \peq$ would be 
(in $k_{\rm B}T$ units, up to a constant)
\begin{align} \label{eq:Phi}
\Phi &= \log(\ell!r!n!) - (\ell+n)\log(\alpha/\beta)- n \log(\gamma/\mu).
\end{align}
When driven out of equilibrium by a dynamic rate $\alpha(t)$, the network
responds with a non-equilibrium distribution $p(\ell,n|t)$.
Importantly, $\Phi(\ell, n, t)$ remains meaningful as the free energy
associated with equilibrium at the current $\alpha(t)$.

\begin{figure}[tb]
\begin{center}
\includegraphics[width=\columnwidth]{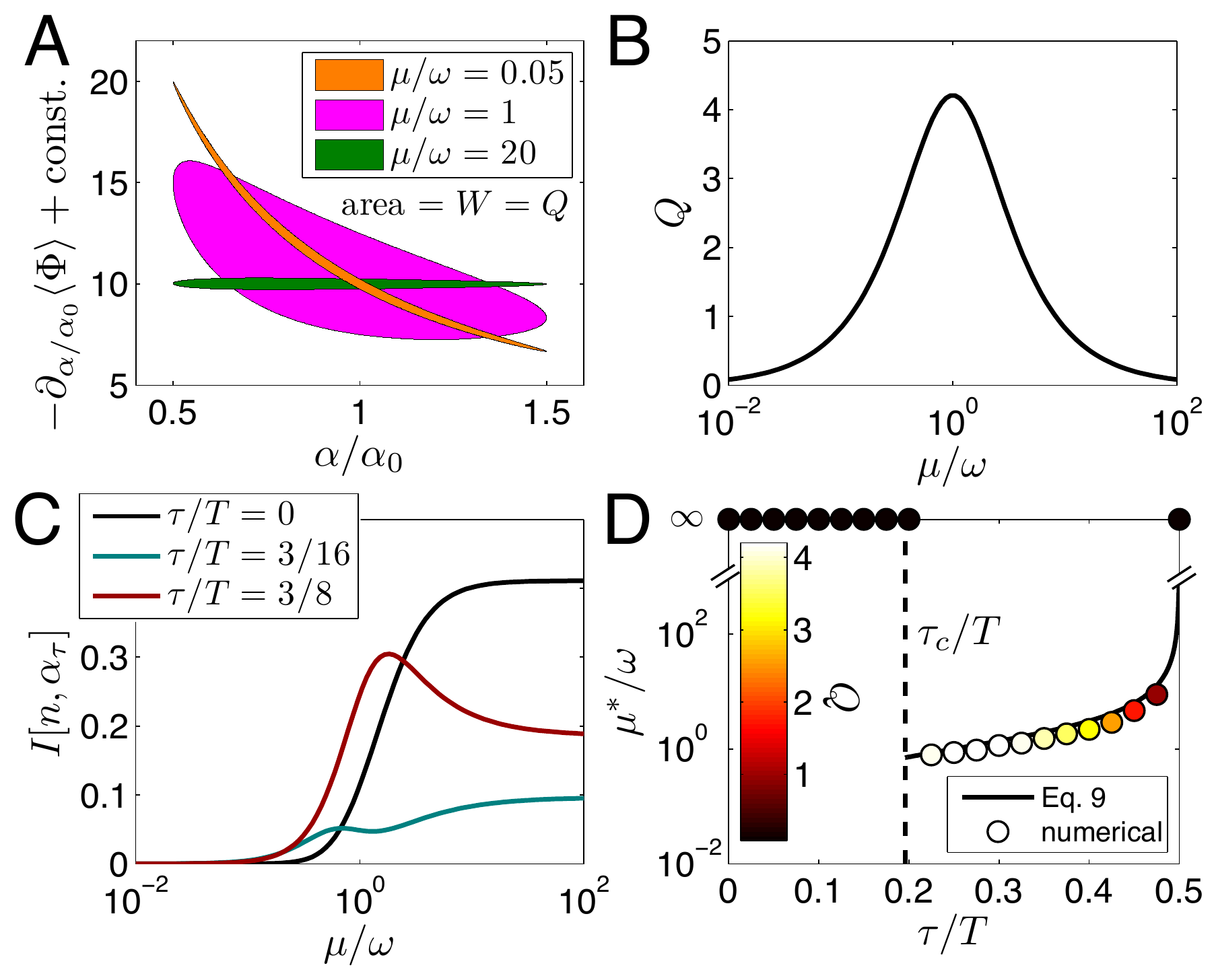}

\caption{Dissipation and prediction in the high copy-number regime. (A)
Dissipated heat is the area of the cycle defined by the thermodynamic force
$-\partial_\alpha\E{\Phi}$ and the driving function $\alpha(t)$. (B) Frequency
matching $\mu = \omega$ maximizes heat. (C) Perfect tracking $\mu\to\infty$
maximizes near-future predictive information (small $\tau$), while frequency
matching $\mu \simeq \omega$ maximizes far-future predictive information (large
$\tau$). (D) Phase diagram of information-optimal $\mu^*/\omega$ vs.\ $\tau$
reveals transition at $\tau_c$, beyond which optimal prediction can be maximally
dissipative. Parameters are $\rho=0.5$ and $\nu_0= 10$.}
\label{fig:work_info}
\end{center}
\end{figure}

After transient relaxation, the system approaches a periodic state, where
\begin{align}
\label{eq:totalderiv}
0 &= \oint dt\, \frac{d}{dt}\E{\Phi}
= \oint dt\,\frac{d}{dt}\sum_{\ell n}p(\ell,n|t)\Phi(\ell,n,t)
\nonumber\\
&= \oint d\alpha(t)\,
\E{\partial_\alpha\Phi} + \oint dt\, \sum_{\ell n} \left
[\partial_t p(\ell,n|t)\right] \Phi(\ell,n,t) \nonumber \\
&\equiv W + (-Q).
\end{align}
Here $\oint\equiv\int_t^{t+T}$ integrates over a period, and $\E{\cdot}$
averages over $p(\ell,n|t)$.
We recognize $W$ as the average chemical work that the external forcing
$\alpha (t)$ performs on the system over the course of a
cycle, Fig.~\ref{fig:work_info}A. We rewrite
$Q$
using
Eq.~\ref{eq:CME} as
\begin{align}
Q
&=\oint dt\, \sum_{\ell n}
\bigl[J_{\ell n}^{\ell_+n}\bigl(-\Delta\Phi_{\ell n}^{\ell_+ n}\bigr) 
+ J_{\ell n}^{\ell_- n_+}\bigl(-\Delta\Phi_{\ell n}^{\ell_- n_+}\bigr)\bigr]
\nonumber\\ 
&\equiv \Qex + \Qbind,
\label{eq:heat} 
\end{align}
where
$J_{\ell n}^{\ell_+ n}=\alpha(t)p(\ell|t)-\beta\ell_+ p(\ell_+|t)$ and $J_{\ell
n}^ {\ell_- n_+}=\gamma\ell r p(\ell,n|t)-\mu n_+p(\ell_-, n_+|t)$
are the net probability fluxes for ligand exchange and ligand binding,
respectively. Eq.~\ref{eq:heat} allocates the dissipated heat per cycle to
free-energy drops $-\Delta\Phi^{n'\ell'}_{n\ell} = \Phi(n,\ell,t)-\Phi
(n',\ell',t)$ occurring in the reaction events. Eq.~\ref{eq:totalderiv} thus
states the first law: The net applied work $W$ over a cycle is dissipated into
the thermal bath as heat $Q$ by the reactions.

We assume throughout that ligand exchange fluxes dominate over binding 
fluxes, i.e.~binding is effectively reaction limited. We thus require
$\beta\gg\gamma \nu_0$ and $\alpha_0\gg\mu\nu_0$ \cite{supplementary}, where
$\nu_0$ is the time-average of $n(t)$. Then $\ell(t)$ is unperturbed by the
receptor state: $\partial_t p(\ell|t) \simeq \B_\ell^{\alpha(t),\beta}
p(\ell|t)$.
We further suppose that ligand numbers respond instantaneously to the input,
$\beta\gg\omega$. Under these two assumptions, $\ell$ is Poisson distributed
with mean $\E{\ell(t)}=\alpha(t)/\beta$,  and ligand exchange is non-dissipative, 
$\Qex=0$ \cite{supplementary}. Evaluating $W$ using Eq.~\ref{eq:Phi},
the energy balance Eq.~\ref{eq:totalderiv} simplifies to
\begin{equation}
\label{eq:Wtrack}
-\oint d\alpha(t)\,\E{n}/\alpha = W =  \Qbind.
\end{equation}

\paragraph{Predictive information}
To assess the sensory performance of the module, we ask a biologically
motivated question: At time $t$, how much does the output $n=n(t)$
enable the cell to prepare for the future environmental state
$\alpha_\tau\equiv\alpha(t+\tau)$?
The answer is given by the mutual
information (in nats),
\begin{equation}
\label{eq:defI}
 I[n,\alpha_\tau] = 
\int d\alpha_\tau \sum_n p(n,\alpha_\tau)
\log\Bigl[\frac{p(n,\alpha_\tau)}{p(n)p(\alpha_\tau)}\Bigr],
\end{equation}
which is the reduction in uncertainty about $\alpha_\tau$, given $n$
\footnote{We note that \eref{defI} is distinct from both the predictive
information between entire past and future trajectories \cite{bialek01} and the
information rate between input and output trajectories \cite{tostevin09};
instead, \eref{defI} focuses on a particular time point shifted by $\tau$ into
the future.}. Here $p(n, \alpha_\tau) = \oint dt\, p[n(t),\alpha(t+\tau)|t]
p(t)$, and $p(t)=1/T$ corresponds to picking a sampling time at random, or
equivalently to sensing a signal of unknown phase.

\paragraph{High copy-number limit}

To gain intuition, we first consider the effect of varying response speed
$\mu/\omega$ on dissipation and predictive information in a high copy-number
limit \cite{supplementary}. Specifically, we let $\{\lambda_0, \nmax\}\to\infty$
and $\gamma\to 0$ such that the mean number of bound receptors $\nu_0 =
\gamma\lambda_0 N/\mu$ remains constant.
The solution $p(n|t)$ is known to be a Poisson distribution \cite{gardiner04,
Mugler2010} with a mean $\E{n(t)}
= \nu_0\{1+\rho\delta\cos[\omega( t-\Lambda)]\}$ that lags behind the driving by
$\Lambda \equiv \tan^{-1}(\omega/\mu)/\omega$ and is damped by a factor $\delta
\equiv [1+(\omega/\mu)^2]^{-1/2}$. 


In this limit, we compute the total heat dissipation by using
$\E{n(t)}$ with Eq.~\ref{eq:Wtrack}, as
\begin{align}
\label{eq:Whighligand}
Q = 2\pi\bigl[1-\sqrt{1-\rho^2}\bigr]\nu_0\frac{\mu/\omega}{1+(\mu/\omega)^2},
\end{align}
shown in Fig.~\ref{fig:work_info}B. Maximum heat dissipation occurs for
frequency-matched $\mu=\omega$, at lag $\Lambda=T/8$. Fast $\mu/\omega\to\infty$
allows instant tracking of the input, dissipating no heat; while slow
$\mu/\omega\to0$ produces no response, also dissipating no heat.

\paragraph{Dissipative optimal prediction} 

The predictive information, evaluated numerically using the known Poissonian
$p(n|t)$ \cite{supplementary}, is shown as a function of the receptor speed
$\mu/\omega$ in Fig.~\ref{fig:work_info}C.  Its shape depends strongly on the
desired prediction interval $\tau$. As may be expected, instantaneous sensing
($\tau=0$) works best when receptor binding tracks the input at
$\mu/\omega\to\infty$, without dissipation. Surprisingly, for finite
$\tau=3T/8$, the prediction performance has a pronounced optimum at a
frequency-matched response with $\mu\simeq\omega$. Thus, long-term prediction
can be optimal at maximum dissipation. This is our principal observation.

Furthermore, the phase diagram Fig.~\ref{fig:work_info}D shows that the optimal
prediction strategy switches from non-dissipative $\mu^*=\infty$ to maximally
dissipative $\mu^*\simeq\omega$, as the forecast interval exceeds a critical
value $\tau_c$.  The discontinuity arises from a secondary local maximum in
$I$ that develops for increasing $\tau$ 
(cf.~Fig.~\ref{fig:work_info}C) and exceeds the plateau value at $\mu\to\infty$
when $\tau\geq\tau_c$. 
Expanding $p(n|t)$ in Fourier modes in $t$ and its natural eigenmodes in $n$
\cite{supplementary}, we obtain
\begin{equation} 
\label{eq:Ismalldriving} 
I[n,\alpha_\tau] =
    \frac{\rho^2\nu_0}{4} \Bigl[ \frac{( \mu/\omega )^2\cos(\omega\tau) -
    \mu/\omega\sin(\omega\tau)}{1+( \mu/\omega )^2} \Bigr]^2 
\end{equation} 
to leading order in the driving amplitude. This expression indeed has a local
maximum at $\mu^*/\omega = \sin(\omega\tau)/ [\cos(\omega\tau)+1]$ that becomes
global when $\tau>\tau_c=T\cos^{-1}(1/3)/(2\pi)\simeq 0.2T$ (\fref{work_info}D).

A second surprising feature of \fref{work_info}C is that in the
frequency-matched regime $\mu\simeq\omega$, the predictive information at $\tau
= 3T/8$ is higher than the instantaneous information at $\tau = 0$.
Counterintuitively, the system can contain more information about the future
than about the present.

\begin{figure}[tb]
\begin{center}
  \includegraphics[width=\columnwidth]{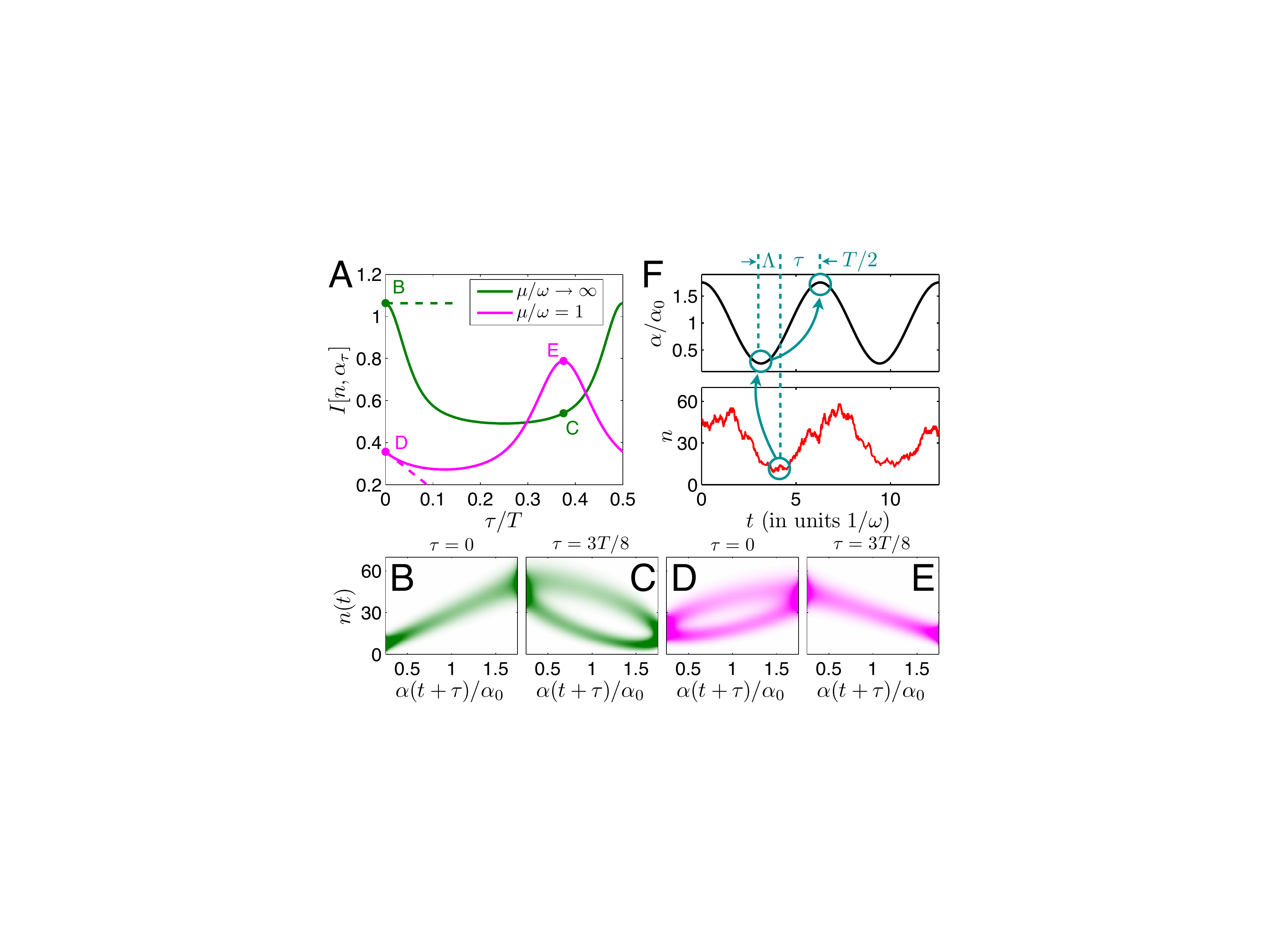}
  \caption{Prediction through memory. 
  (A) For a range of prediction intervals $\tau$, frequency matching
  $\mu=\omega$ is more predictive than perfect tracking $\mu\to\infty$. 
  The reason, as demonstrated by $p(n,\alpha_\tau)$ at key points (B-E as
  indicated in A) and illustrated in F, is memory: receptors with $\mu=\omega$ 
  time-integrate the signal, so that $n(t)$ lags by $\Lambda$; it is maximally
  correlated with $\alpha(t-\Lambda)$, which in turn mirrors 
  $\alpha(t-\Lambda+T/2)$.
  Parameters are $\rho = 0.75$ and $\nu_0 = 30$. 
  }
  \label{fig:explanation}
\end{center}
\end{figure}

\paragraph{Exploiting signal correlations through memory} 

The above observations can be understood by carefully considering the connection
between prediction and memory. \fref{explanation}A shows the predictive
information $I[n,\alpha_\tau]$ as a function of the prediction interval, denoted
$I_\infty(\tau)$ and $I_\omega(\tau)$, for fast ($\mu/\omega\to\infty$) and
frequency-matched ($\mu = \omega$) response, respectively. Three features are
apparent at small $\tau$. First, $I_\infty(0)$ is larger than $I_\omega(0)$,
i.e.~responding quickly does maximize information about the present. Second,
both $I_\infty(\tau)$ and $I_\omega(\tau)$ decrease for small $\tau$, in line
with a general expectation that longer-term prediction is more difficult.  %
Third, as illustrated by the slopes in \fref{explanation}A (dashed lines),
$\partial_\tau I_\omega\bigr|_0$ is steeper than $\partial_\tau
I_\infty\bigr|_0$, which is zero. In fact, differentiating \eref{Ismalldriving}
with respect to $\tau$, taking $\tau\to 0$, and comparing to \eref{Whighligand},
we obtain $-\partial_\tau I\bigr|_0 \le Q/T$ \cite{supplementary}, which is
precisely the bound of Ref.~\cite{still12a} applied to our system.

The key point is that a surprise occurs beyond the limit $\tau\to 0$.
Both curves increase again, which is a direct consequence of
the sinusoidal input: When $\tau$ is a half-integer or integer multiple of $T$,
$I(\tau)$ must equal $I(0)$, since then $\alpha(t+\tau)$ perfectly
mirrors or tracks $\alpha(t)$, respectively.  Indeed, $I(\tau)$ is
$T/2$-periodic.
Interestingly, the frequency-matched $I_\omega(\tau)$ increases sufficiently to
overtake both its own initial value, and the fast-responding $I_\infty(\tau)$. 

The takeover occurs because the lag introduced by a slower response 
removes an ambiguity inherent in prediction, \fref{explanation}B-E.  While the
fast response tracks the present input well (B), its predictions suffer from a
two-fold ambiguity about $\alpha(t+\tau)$, since a given value of $n(t)$ maps
with high probability to two distinct values of $\alpha(t+\tau)$ (C),
corresponding to the rising and falling half-period.
In contrast, the frequency-matched response tracks the delayed signal
$\alpha(t-\Lambda)$.  This introduces a two-fold ambiguity about the present
signal for the same reason (D) but strikingly, helps prediction: The future
signal $\alpha(t-\Lambda+T/2)$ is tracked without two-fold ambiguity (E).
Indeed Fig.~\ref{fig:explanation}A shows that $I_\omega(\tau)$ is maximal when
the lag, advanced by half a period, equals the prediction interval: $-\Lambda +
T/2=\tau$, or, for $\Lambda=T/8$ in the maximally dissipative case, when
$\tau=3T/8$. This advantage of removing ambiguity outweighs the disadvantage of
a reduced response range due to damping; the net effect is an increase in
predictive power over the fast response.

In essence, a dissipative system can exploit memory to achieve superior
predictive power, by maximizing information of the past and therefore, due to
strong signal autocorrelations, about the future (Fig.~\ref{fig:explanation}F).


\begin{figure}[tb]
\begin{center}
  \includegraphics[width=\columnwidth]{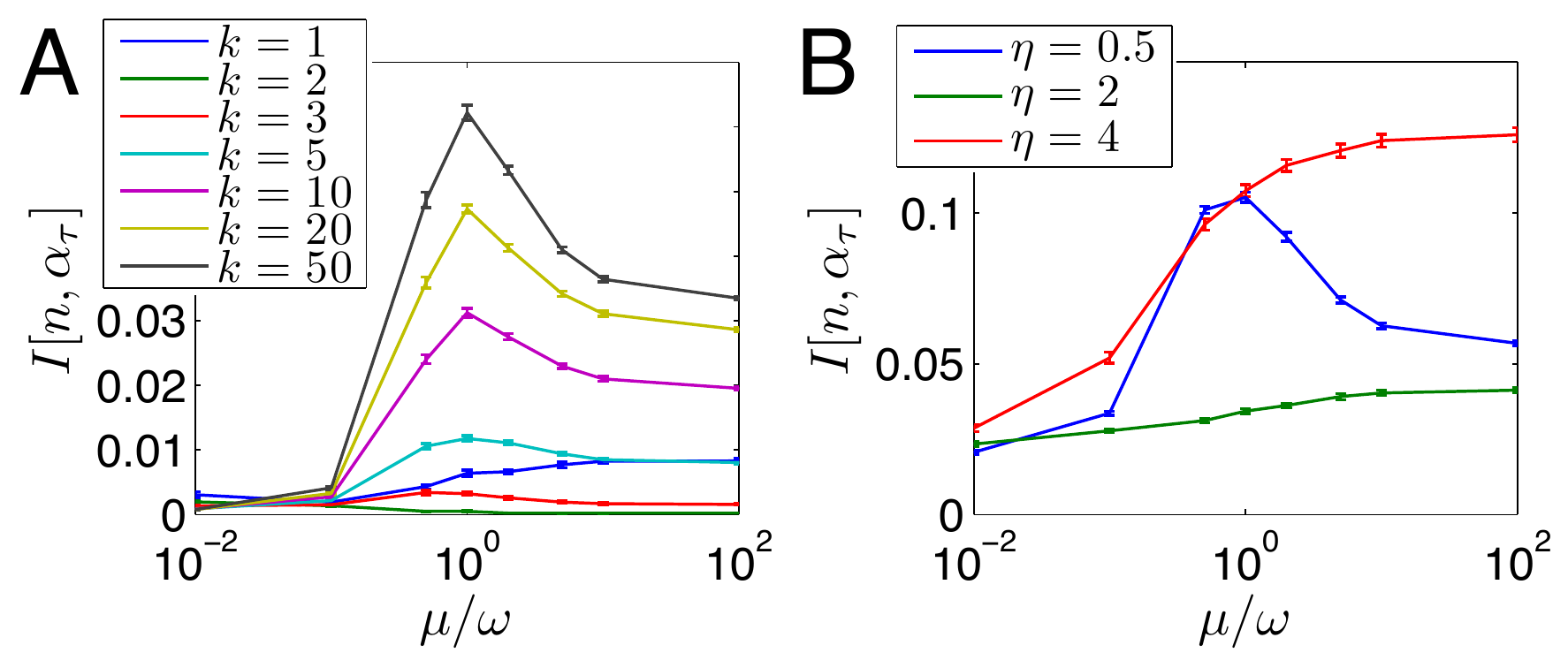}
  \caption{Dissipative optimal prediction is generic. 
  (A) For stochastic two-state driving with
  Gamma-distributed waiting times, dissipative prediction is optimal for shape 
parameters $k\ge 3$.
(B) For driving by a non-Markovian damped
  harmonic oscillator, dissipative prediction is optimal in the underdamped
  regime $\eta=0.5$ but not at critical
  damping $\eta=2$ or in the overdamped regime $\eta=4$. 
  Parameters are $\tau = 3T/8$, $\rho=0.5$, $\lambda_0=25$, $N=25$,
  $\beta/\omega=100$, and
  $\gamma=\mu/\lambda_0$. Simulation details are in \cite{supplementary}.}
\label{fig:generic}
\end{center}
\end{figure}

\paragraph{Dissipative optimal prediction is generic} 

While tractable, a sinusoidal signal is arguably too predictable.  In fact,
since $I(\tau)$ is periodic, the optimal prediction strategy alternates
indefinitely between dissipative and non-dissipative as $\tau\to\infty$.
However, dissipative optimal prediction occurs in more general contexts; we now
explore what features of signal and response are necessary to observe it. 

The input distribution $p(\alpha)$ for a sinusoid is peaked at the extrema, but
this is not necessary: A triangle wave with uniform $p(\alpha)$ also produces
the effect (not shown). High copy number is not necessary: Relaxing the high
copy-number limit, we observe dissipative optimal prediction even down to
$\lambda_0 = 1$ and $N = 1$ \cite{supplementary}.


Periodic and deterministic signals are not necessary. This is seen by
considering two-state driving processes that switch between $\alpha_0(1\pm
\rho)$ with waiting times that are Gamma-distributed with varying shape
parameter but constant mean $T/2=\pi\omega$. Fig.~\ref{fig:generic}A shows that
dissipative prediction is optimal for strongly correlated two-state signals, and
even for very stochastic signals down to shape parameter $k\geq 3$, but not for
the Markovian random telegraph signal at $k=1$.  Indeed, we find that prediction
of Markovian driving signals is governed purely by the instantaneous response
$p[n(t)|\alpha(t)]$, as no additional predictive information is encoded in past
signal values \cite{supplementary}.

While a non-Markovian signal appears necessary, it may be insufficient. This is
seen by considering a continuous, non-Markovian driving process $\alpha(t) =
\alpha_0[1+\rho x(t)]$ defined by the two-dimensional Langevin equation for a
thermal harmonic oscillator, $\partial_{\omega t} x = p,\;  \partial_{\omega
t}p = -x -\eta p + \sqrt{2\eta} \xi,$ where $\xi$ is unit Gaussian white noise.
This signal does not admit dissipative optimal prediction in the overdamped
regime $\eta\geq\eta_c=2$ (Fig.~\ref{fig:generic}B). Only when the driving
becomes oscillatory with resonant frequency $\omega_1 =
\omega\sqrt{1-\eta^2/4}>0$ for $\eta<\eta_c$, is dissipative optimal prediction
recovered.

\paragraph{Discussion} 

Our model system Eq.~\ref{eq:reactions} shows that dissipation is not 
fundamentally required for measuring the current state of a signal. In fact,
dissipation degrades instantaneous sensing by introducing ambiguity and damping
the response. However, a lagging, dissipative response improves finite-time
prediction by resolving ambiguities in non-Markovian signals, over a range of
prediction intervals on the order of the signal period.

This may be a general mechanism for biological systems to anticipate
oscillations. In cyanobacteria, the KaiABC circadian clock \cite{johnson11}  is
read out by binding of the effector kinase SasA to the oscillating
phosphorylated form of KaiC. The low dissociation rate $\mu\simeq 4\times 10^
{-4}/$s \cite{valencia12} of this complex implies a lagging response with $\mu/
\omega\simeq 5$, enabling dissipative prediction.
In sea urchin sperm cells, chemotaxis is performed by helical swimming up a
ligand gradient, giving rise to an oscillatory ligand signal with period $T
\simeq 1$ s \cite {kaupp08}; as in \eref{reactions},
ligands bind and activate transmembrane receptors whose deactivation rate $\mu/
\omega\simeq 0.5$ also enables dissipative prediction. Intriguingly, each
ligand peak is eventually followed by a motor response after $\tau=0.4T$
\cite{kaupp08}, such that the lagging receptor state would optimally predict
the signal at the future time $t+\tau$ of the motor response (cf.~Fig.~\ref
{fig:explanation}).

In this study, energy is supplied by a dynamic input and dissipated by a
passive sensory module.  Many signal transduction modules dissipate energy that
is supplied internally by ATP turnover.  It will be interesting to explore the
interplay between input-supplied and internally supplied energy and its
implications for sensing and prediction. 


This work is part of the research program of the ``Stichting FOM'', which is
financially supported by the ``Nederlandse organisatie voor Wetenschappelijk
Onderzoek'' (NWO). NB was supported in part by a fellowship of the Heidelberg
Center for Modelling and Simulation in the Biosciences (BIOMS).

%



\clearpage

\appendix

\renewcommand{\theequation}{A\arabic{equation}}
\setcounter{equation}{0}
\renewcommand{\thefigure}{A\arabic{figure}}
\setcounter{figure}{0}

\setcounter{page}{1}
\renewcommand{\thepage}{A\arabic{page}}


\appendix
\section{Appendix}

\subsection{Dominant ligand fluxes}

To identify the conditions for which ligand fluxes dominate over binding fluxes, 
we sum Eq.~\ref{eq:CME} over $n$, obtaining (here suppressing $t$ arguments) 
\begin{multline}\label{eq:ellmarginal}
\dot p(\ell) = \B_\ell^{\alpha,\beta} p(\ell) 
+ \gamma[\E{r|\ell_+}\ell_+p(\ell_+) - \E{r|\ell}\ell p(\ell)] \\
+ \mu[\E{n|\ell_-}p(\ell_-) - \E{n|\ell}p(\ell)].
\end{multline}
For moderate driving amplitude around half-filling of the receptors, the
conditional averages $\E{n|\ell}, \E{r|\ell}$ remain close to their mean
$\nu_0$. Therefore, taking $\beta\gg\gamma \nu_0$ and $\alpha_0\gg\mu\nu_0$
ensures that $\dot p(\ell)\simeq \B_\ell p(\ell)$, ie.~the ligand in- and
outflux terms dominate.

To see explicitly that the ligand exchange heat vanishes for fast ligand
exchange, we rewrite
\begin{align}\label{eq:Qexvanish}
\Qex 
&= \oint\sum_{\ell}[\alpha(t)p(\ell|t)-\beta\ell_+ p(\ell_+|t)]
\log\Bigl[\frac{\alpha(t)}{\beta\ell_+}\Bigr]\,dt \nonumber\\
&= \oint \bigl(\alpha - \beta \E{\ell}\bigr)\log(\alpha/\beta) \,dt\nonumber\\
&\quad -\oint \sum_{\ell} \bigl[\alpha p(\ell|t) - \beta\ell_+
p(\ell_+|t)\bigr]\log\ell_+ \,dt.
\end{align}
For fast ligand exchange, $p(\ell|t)$ is a Poisson
distribution with mean $\E{\ell(t)}=\alpha(t)/\beta$, so that the first
integral on the right hand side of the last equality vanishes. To see that the
second integral vanishes, note that the Poisson distribution satisfies
$p(\ell_+)/p(\ell)=\E{\ell}/\ell_+$.

\subsection{High copy-number limit}

We let $\{\lambda_0, \nmax\}\to\infty$ and $\gamma\to 0$ such that the mean
number of bound receptors $\nu_0 = \gamma\lambda_0 N/\mu$ remains constant.
This ensures that $n(t)$ remains in the linear, non-saturated range of the
response curve: $\nmax\gg\nu_0$ implies that $r(t)=O(\nmax - \nu_0) =
O(\nmax)$. It also ensures that the receptors are effectively driven by a
deterministic signal: The relative width of the ligand distribution decreases
as $\sigma_\ell/\lambda_0= O(\lambda_0)^{-1/2}$.  The receptor dynamics thus
reduce to a birth-death process, $\partial_t p(n|t) = \B_n^{\g(t), \mu}p(n|t)$,
with effective birth rate $\gamma\ell(t) r(t) = \gamma [\alpha(t)/\beta]N +
O(\nu_0/\nmax) + O(\lambda_0^{-1/2})$, which we denote as $\g(t)\equiv \gamma
[\alpha(t)/\beta]N = \nu_0\mu[1+\rho\cos(\omega t)]$. The solution to a
birth-death process with time-dependent birth rate $\g(t)$ is $p(n|t) =
\int_{-\infty}^0 \,dt' e^{-\mu(t-t')} \g(t') = \nu_0\{1+\rho\delta\cos[\omega(
t-\Lambda)]\}$, with lag $\Lambda \equiv \tan^{-1}(\omega/\mu)/\omega$ and
damping $\delta \equiv [1+(\omega/\mu)^2]^{-1/2}$, as in the main text.

\subsection{Analytic results for predictive information}

For a deterministic signal $\alpha(t)$, we have $p(\alpha_\tau|n,t)=p(\alpha_
\tau|t)=\delta[\alpha_\tau-\alpha(t+\tau)]$, and the predictive information 
(\eref{defI}) becomes
\begin{align}
 I[n,\alpha_\tau]
=& \sum_n \int d\alpha_\tau\ p(n,\alpha_\tau)\nonumber\\
=& \sum_n \int d\alpha_\tau \oint dt\ p(\alpha_\tau|n,t)p(n|t)p(t)
\log\Bigl[\frac{p(n|\alpha_\tau)}{p(n)}\Bigr]\nonumber\\
=& \sum_n \frac{1}{T} \oint dt\ p(n|t)
\log\Bigl[\frac{p(n|\alpha(t+\tau))}{p(n)}\Bigr].
\end{align}
For cosine driving $\alpha(t)=\alpha(T-t)$, there is a two-to-one relationship 
between $t$ and $\alpha$.  This yields $p(n|\alpha(t+\tau))=[p(n|t_1)+p(n|t_2)]/
2$, where $t_1=t$ and $t_2=T-t-2\tau$ are the two time points for which $
\alpha_\tau$ takes on the value $\alpha(t+\tau)$.  The predictive information 
becomes
\beq
\elabel{Isimp}
 I[n,\alpha_\tau]
= \sum_n \frac{1}{T} \oint dt\ p(n|t)
\log\Bigl[\frac{p(n|t_1)+p(n|t_2)}{2p(n)}\Bigr].
\eeq
In the high copy-number limit (Figs.\ \ref{fig:work_info} and \ref
{fig:explanation}), \eref{Isimp} is evaluated numerically using the Poissonian 
$p(n|t)$ with mean $\E{n(t)} = \nu_0\{1+\rho\delta\cos[\omega( t-\Lambda)]\}$,
time lag $\Lambda \equiv \tan^{-1}(\omega/\mu)/\omega$, and damping factor $
\delta \equiv [1+(\omega/\mu)^2]^{-1/2}$.

To get analytical insight, we can expand \eref{Isimp} in the limit of small
driving amplitude $\rho$. To facilitate the expansion, we exploit the fact that
$p(n|t) $ can be expressed \cite{Mugler2010} in terms of its Fourier modes in
$t$,
\beq
\elabel{FT}
p(n|t) = \sum_{z=-\infty}^\infty p_n^z e^{-iz\omega t},
\eeq
and its natural eigenmodes in $n$,
\beq
\elabel{pnz}
p_n^z = e^{iz\omega\Lambda}\sum_{j=0}^\infty
	\frac{(\nu_0\rho\delta/2)^{2j+|z|}}{j!(j+|z|)!}\phi_n^{2j+|z|}.
\eeq
Here $p_n^z=(1/T)\int_0^Tdt\ e^{iz\omega t}p(n|t)$ are the components of the
Fourier transform, which have support only at integer multiples $z$ of the
driving frequency, and $\phi_n^j$ are the eigenmodes of the static birth-death
process with mean bound receptor number $\nu_0$, i.e. $-\B_n^{\nu_0,1}\phi_n^j=j
\phi_n^j$ for eigenvalues $j\in\{0,1,\dots,\infty\}$ \cite{Walczak2009}.
\eref{pnz} shows directly that the distribution is expressible as an expansion
in the small parameter $\rho$. The remaining task is then to identify the
leading term in $\rho$.

To identify the leading term in $\rho$, we insert \eref{FT} into \eref{Isimp}, 
which yields
\beqn
\elabel{I5}
I &=& \frac{1}{T}\sum_n\int_0^T dt\ \sum_z p_n^z e^{-iz\omega t}\nonumber\\
&&	\times\log \left\{ \frac{1}{2p_n^0}\sum_{z'}p_n^{z'}
	\left[e^{-iz'\omega t}+e^{-iz'\omega (T-2\tau-t)}\right]\right\}.\qquad
\eeqn
Here we have recognized that $p(n)$, which is the time average of $p(n|t)$, is 
also the zeroth Fourier mode:
$p(n)=\int_0^Tdt\ p(n|t)p(t)=(1/T)\int_0^Tdt\ p(n|t)=p_n^0$.
Isolating the $z'=0$ term and defining $q_n^z\equiv p_n^ze^{iz\omega\tau}$ to 
make the expression more symmetric yields
\beqn
I &=& \frac{1}{T}\sum_n\int_0^T dt\ \sum_z p_n^z e^{-iz\omega t}\nonumber\\
\elabel{I6}
&&	\times\log \left\{1+\frac{1}{2p_n^0}\sum_{z'\neq0}q_n^{z'}
	\left[e^{-iz'\omega(t+\tau)}+e^{iz'\omega(t+\tau)}\right]\right\},\qquad
\eeqn
where we have recognized that $e^{-iz'\omega T}=1$.  Then, recognizing that the 
term in brackets is symmetric upon $z'\to-z'$, we write the $z'$ sum in 
terms of only positive integers,
\beqn
I &=& \frac{1}{T}\sum_n\int_0^T dt\ \sum_z p_n^z e^{-iz\omega t}\nonumber\\
\elabel{I7}
&&	\times\log \left\{1+\frac{1}{2p_n^0}\sum_{z'>0}r_n^{z'}
	\left[e^{-iz'\omega(t+\tau)}+e^{iz'\omega(t+\tau)}\right]\right\},\qquad
\eeqn
where
\beqn
r_n^z &\equiv& q_n^z+q_n^{-z}\nonumber\\
\elabel{rp}
&=& e^{iz\omega\tau}p_n^z+e^{-iz\omega\tau}p_n^{-z}\nonumber\\
&=& \left[e^{iz\omega(\Lambda+\tau)}+e^{-iz\omega(\Lambda+\tau)}\right]
	\sum_j \frac{(\nu_0\rho\delta/2)^{2j+|z|}}{j!(j+|z|)!}\phi_n^{2j+|z|}
\nonumber\\
\elabel{rnz}
&=& 2\cos[z\omega(\Lambda+\tau)]
	\sum_j \frac{(\nu_0\rho\delta/2)^{2j+|z|}}{j!(j+|z|)!}\phi_n^{2j+|z|}
\eeqn
is a real quantity.

Now, since $r_n^z$ is expressed in terms of our small 
parameter $\rho$, we Taylor expand the log in \eref{I7}:
\beqn
I &=& \frac{1}{T}\sum_n\int_0^T dt\ \sum_z p_n^z e^{-iz\omega t}
	\sum_{k=1}^\infty\frac{(-1)^{k+1}}{k}\nonumber\\
\elabel{I8}
&&	\times\left\{\frac{1}{2p_n^0}\sum_{z'>0}r_n^{z'}
	\left[e^{-iz'\omega(t+\tau)}+e^{iz'\omega(t+\tau)}\right]\right\}^k.\quad
\eeqn
It will turn out that the first two terms in the Taylor expansion will 
contribute to the leading order in $\rho$.  The first term ($k=1$) is
\beqn
I^{(1)} &=& \sum_n	\frac{1}{2p_n^0}\sum_{z'>0}r_n^{z'}
	\sum_z p_n^z\nonumber\\
\elabel{I9}
&&	\times\frac{1}{T}\int_0^T dt\ e^{-iz\omega t}
	\left[e^{-iz'\omega(t+\tau)}+e^{iz'\omega(t+\tau)}\right],\qquad
\eeqn
where we have reordered terms in preparation for exploiting the relation
$(1/T)\int_0^T dt\ e^{-i(z-z')\omega t}=\delta_{zz'}$.  This relation turns the 
two terms in brackets into Kronecker deltas, which each collapse the sum over 
$z$, leaving
\beqn
I^{(1)} &=& \sum_n \frac{1}{2p_n^0}\sum_{z'>0}r_n^{z'}
	\left(e^{-iz'\omega\tau}p_n^{-z'}+e^{iz'\omega\tau}p_n^{z'}\right)
	\nonumber\\
\elabel{I11}
&=& \frac{1}{2}\sum_n\frac{1}{p_n^0}\sum_{z'>0}(r_n^{z'})^2.
\eeqn
In a completely analogous way, the second term in the Taylor expansion ($k=2$) 
reduces to
\beq
\elabel{I12}
I^{(2)} = -\frac{1}{8}\sum_n \frac{1}{\left(p_n^0\right)^2}
	\sum_{x>0}\sum_{y>0}r_n^{x}r_n^{y}
	\left(r_n^{x+y}+r_n^{x-y}\right).
\eeq
Considering the $j=0$ term in $r_n^z$ (\eref{rnz}), it is clear that the leading 
order behavior in $\rho$, proportional to $\rho^2$, comes from the $z'=1$ term 
in \eref{I11} and the $x=y=1$ term in \eref{I12}:
\beqn
\elabel{I13}
I &\approx& \frac{1}{2}\sum_n \frac{1}{p_n^0}(r_n^1)^2
	-\frac{1}{8}\sum_n \frac{1}{\left(p_n^0\right)^2}
	r_n^{1}r_n^{1}\left(r_n^{0}\right)\\
\elabel{I14}
&=& \frac{1}{2}\sum_n \frac{(r_n^1)^2}{r_n^0}\\
\elabel{I15}
&\approx& \cos^2[\omega(\Lambda+\tau)]
	\left(\frac{\nu_0\rho\delta}{2}\right)^2
	\sum_n\frac{(\phi_n^1)^2}{\phi_n^0}\\
\elabel{I16}
&=& \frac{\nu_0^2\rho^2}{4}
	\left[ \frac{\cos(\omega\tau) - (\omega/\mu)\sin(\omega\tau)}
	{1+(\omega/\mu)^2} \right]^2
	\sum_n\frac{(\phi_n^1)^2}{\phi_n^0}.\qquad
\eeqn
Here, \eref{I14} uses the fact that $r_n^0=2p_n^0$ (\eref{rp}), \eref{I15} takes 
only the $j=0$ term in \eref{rnz}, and \eref{I16} recalls that $\delta = [1+
(\omega/\mu)^2]^{-1/2}$ and uses
\beqn
\cos[\omega(\Lambda+\tau)]
&=& \cos(\omega\Lambda)\cos(\omega\tau)-\sin(\omega\Lambda)\sin(\omega\tau) 
\nonumber\\
&=& \cos\left[\tan^{-1}(\omega/\mu)\right]\cos(\omega\tau)\nonumber\\
&&	-\sin\left[\tan^{-1}(\omega/\mu)\right]\sin(\omega\tau)\nonumber\\
&=& \frac{1}{\sqrt{1+(\omega/\mu)^2}}\cos(\omega\tau)\nonumber\\
\elabel{cosprop}
&&	-\frac{\omega/\mu}{\sqrt{1+(\omega/\mu)^2}}\sin(\omega\tau).
\eeqn
The sum in \eref{I16} is evaluated by noting that the zeroth eigenmode is a 
Poisson distribution with mean $\nu_0$ and that the first eigenmode is related 
to the zeroth eigenmode via $\phi_n^1 = \phi_{n-1}^0-\phi_n^0 = \phi_n^0(n-
\nu_0)/\nu_0$ \cite{Mugler2009}.  The sum therefore becomes $(1/\nu_0^2)\sum_n
\phi_n^0(n-\nu_0)^2$, which is the variance of the Poisson distribution (equal 
to $\nu_0$) divided by $\nu_0^2$, or $1/\nu_0$.  Altogether, then, \eref{I16} 
becomes
\beq
\elabel{I17}
I = \frac{\nu_0\rho^2}{4}
	\left[ \frac{\cos(\omega\tau) - (\omega/\mu)\sin(\omega\tau)}
	{1+(\omega/\mu)^2} \right]^2,
\eeq
as in \eref{Ismalldriving}.

\subsection{Dissipation bounds instantaneous prediction}
In the high copy-number and small-amplitude driving limits, $I$ is given by 
\eref{I17}.  Differentiating \eref{I17} with respect to $\tau$ and evaluating at 
$\tau = 0$ obtains
\beq
\elabel{Icrooks}
\partial_\tau I\big|_0 = \frac{\nu_0\rho^2\omega}{2}
\frac{\omega/\mu}{[1+(\omega/\mu)^2]^2}.
\eeq
Dissipated heat in the high copy-number limit is given by \eref{Whighligand}.  
In the small-amplitude limit, we take \eref{Whighligand} to leading order in $
\rho$:
\beq
\elabel{Qcrooks}
Q = \pi\rho^2\nu_0 \frac{\mu\omega}{\mu^2+\omega^2}.
\eeq
Using $T=2\pi/\omega$, \erefs{Icrooks}{Qcrooks} satisfy
\begin{equation}
\label{eq:stillcrooks}
-\partial_\tau I\big|_0 = \frac{Q/T}{1+(\omega/\mu)^2}.
\end{equation}
Since $[1+(\omega/\mu)^2]^{-1} \le 1$ always, we obtain
$-\partial_\tau I|_0 \le Q/T$, i.e.\ the dissipation bounds the instantaneous 
rate of change of the predictive information, as in \cite{still12a}.

\subsection{Robustness to low copy-number effects}

We here relax the assumption of high copy number and solve numerically the full 
description of the system given by \eref{CME}.  We find that dissipative 
prediction remains optimal as the mean ligand number $\lambda_0$ and the total 
receptor number $N$ are reduced, even down to $\lambda_0 = 1$ (\fref
{copynumber}A) and $N = 1$ (\fref{copynumber}B).  As $N$ is reduced, the 
information is reduced for all values of the response rate $\mu$ (B), since 
reducing $N$ compresses the response range.  As $\lambda_0$ is reduced, the 
information is largely unchanged (A); this is because ligand exchange remains 
faster than the driving dynamics ($\beta/\omega\gg 1$), meaning that even a 
small number of ligand molecules can cycle in and out of the system many times 
over a period.  In both cases, there remains an optimum in the predictive 
information as a function of $\mu$ located in the dissipative regime $\mu \simeq 
\omega$, illustrating that dissipative optimal prediction persists even at low 
copy numbers.

\begin{figure}[tb]
\begin{center}
  \includegraphics[width=\columnwidth]{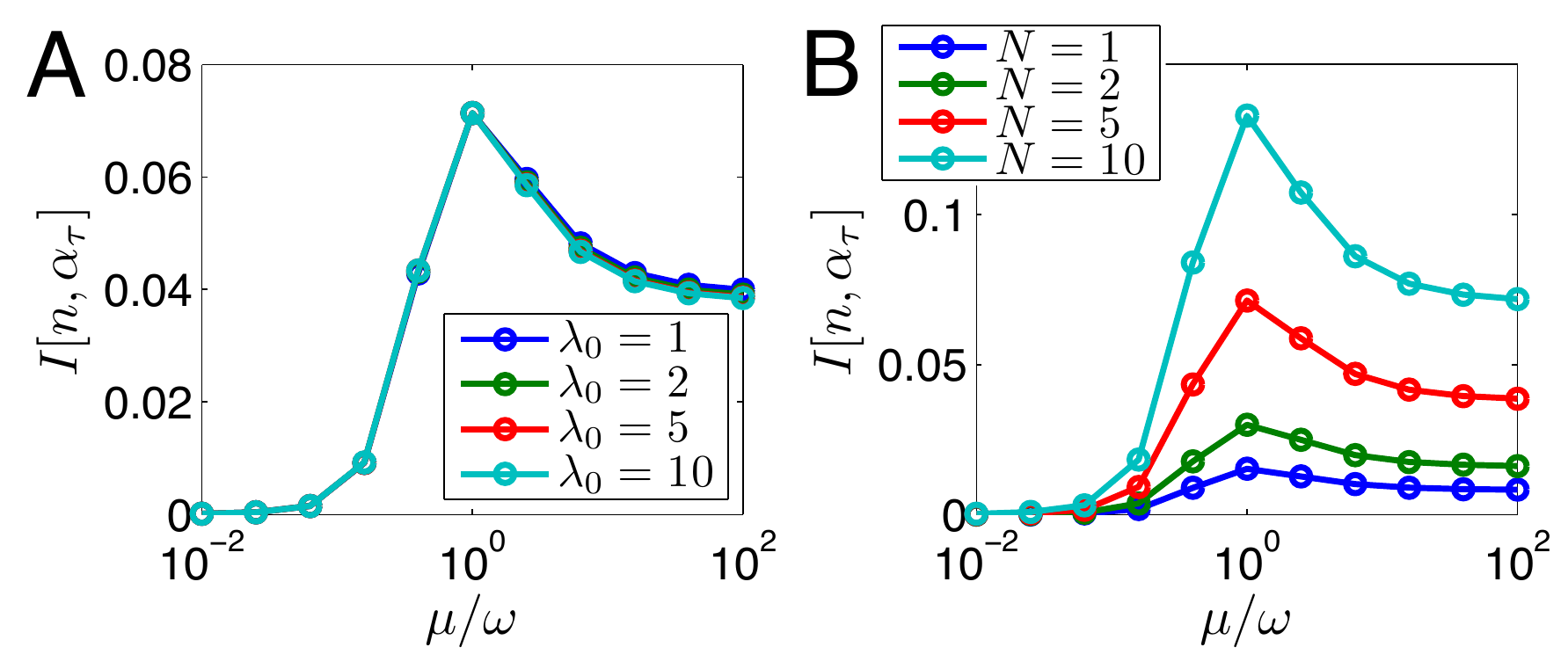}
  \caption{Dissipative optimal prediction persists (A) for low mean ligand 
number $\lambda_0$ and (B) for low total receptor number $N$. Parameters are
$\tau = 3T/8$,
$\rho = 0.5$, $\beta/\omega = 100$, $\gamma = \mu/\lambda_0$, and, in A, $N=5$ 
and, in B, $\lambda_0 = 5$.  In A, all curves closely overlap.}
\label{fig:copynumber}
\end{center}
\end{figure}

\subsection{Dissipative prediction and Markovian driving}

We investigate whether dissipative sensing can in principle improve the
prediction of Markovian inputs. In this section, we will denote the full
signal history up to but excluding $t$ by $\hal$, the present signal by
$\alpha=\alpha(t)$, the present output by $n=n(t)$, and a future signal by
$\alpha_\tau = \alpha(t+\tau)$, respectively. 

For the sensing module studied here, the output depends only on past signal
values and does not feed back onto the signal. This implies that the future
trajectory of the signal does not depend on the particular response of the
output: $p(\alpha_\tau|n, \hal, \alpha, t) = p(\alpha_\tau|\hal, \alpha, t)$. In
other words, the output cannot give any information about the future signal in
excess of what's contained in the signal history. This relation is valid
irrespective of the type of input signal.

We can then rewrite the predictive two-point distribution $
p(\alpha_\tau,n|\alpha, t)$ as an integral over the driving history:
\begin{flalign}\label{eq:jointcond1}
p(\alpha_\tau,n|\alpha, t)
&= \int \mathcal D\hal p(\alpha_\tau, n, \hal|\alpha, t) 
	\nonumber\\
&= \int \mathcal D\hal p(\alpha_\tau| n, \hal,\alpha, t)
	\nonumber\\
& \hphantom{\sum\sum} \times p(n|\hal, \alpha, t)p(\hal|\alpha, t)
	\nonumber\\
&= \int \mathcal D\hal p(\alpha_\tau| \hal,\alpha, t)
	\nonumber\\
& \hphantom{\sum\sum} \times p(n|\hal, \alpha, t)p(\hal|\alpha, t),
\end{flalign}
where the second equality factorizes the integrand using the rule $p(x,y|z) =
p(x|y,z)p(y|z)$, and the last equality uses the no-feedback relation given
above.

If the response is instantaneous (adiabatic), we have the relation $p(n|\hal,
\alpha, t) = p(n|\alpha,t)$; if the input is Markovian, $p(\alpha_\tau|\hal,
\alpha,t) = p(\alpha_\tau|\alpha,t)$. In both cases we can integrate
Eq.~\ref{eq:jointcond1} over $\hal$ trivially and obtain the result:
\begin{equation}\label{eq:jointcond2}
p(\alpha_\tau,n|\alpha, t) =  p(n|\alpha, t)p(\alpha_\tau|\alpha, t).
\end{equation}
When the driving process is stationary, the time dependence in
Eq.~\ref{eq:jointcond2} disappears, and we can write
\begin{align}\label{eq:jointcond3} 
p(\alpha_\tau,n|\alpha) = p(n|\alpha)p(\alpha_\tau|\alpha).
\end{align} 
Thus the basic observation is that for instantaneous response or Markovian
input, $\alpha_\tau$ and $n$ are independent when conditioned on $\alpha$; or
equivalently, the variables $n \leftrightarrow \alpha \leftrightarrow
\alpha_\tau$ form a Markov chain \cite{cover12}. In the following we consider
only Markovian input. We may then insert a further input variable at an
intermediate future time to extend the Markov chain as $n\leftrightarrow \alpha
\leftrightarrow \hat\alpha \leftrightarrow \alpha_\tau$.

The information processing inequality states that mutual information across a
Markov chain is bounded from above by the mutual information across any
subchain. In particular, $I[n,\alpha_\tau]\leq I[n,\hat\alpha]$. By inserting
$\hat\alpha$ at arbitrary intermediate times, this implies that $I(\tau) =
I[n,\alpha_\tau]$ decreases monotonically as a function of the prediction
interval. Thus, for Markovian driving, the output can never contain more
information about the future than about the present signal. This is in contrast
to the non-monotonic behavior shown in Fig.~\ref{fig:explanation}A for a
deterministic signal, and also observed for the noisy non-Markovian signals
we studied (not shown).

The information processing inequality further yields $I[n,\alpha_\tau]\leq
I[\alpha,\alpha_\tau]$: Predictions based on measuring $n$ cannot surpass those
based on the current input $\alpha$. This means that encoding the history of a
Markovian signal by means of a dissipative response has no intrinsic value.
Alternatively, rewriting $p(n,\alpha,\alpha_\tau) = p(\alpha,\alpha_\tau)
p(n|\alpha)$, shows that the properties of the response enter only via the
instantaneous distribution $p(n|\alpha)$, so that any memory is irrelevant for
the prediction performance.

The preceding argument does not prove that a lagging response is always
disadvantageous for Markovian input, but only that it is possible in principle
to construct an instantaneous, non-dissipative system which is as good a
predictor as any given lagging system. Nonetheless, in the cases we studied, a
fast response always produced better predictors of Markovian input signals than
a lagging response.

\subsection{Parameters and details of the simulations}

The data for Figs.~\ref{fig:generic}A and B were generated by a Gillespie-type
kinetic Monte Carlo simulation. Dynamic ligand birth rates $\alpha(t)$ were
approximated as constant during short discretization intervals of length
$\tau_\alpha=T/50$; after each such interval, queued next reaction times were
erased and re-generated according to the new value of the rate. This is an exact
simulation procedure for the approximated system with stepwise-constant rate. We
found that the dissipated heat due to ligand exchange $\Qex$ does depend on
$\tau_\alpha$ and requires much finer rate discretization to vanish (as required
by Eq.~\ref{eq:Qexvanish}). Here, we were interested mainly in $\Qbind$ which
was found to be independent of $\tau_\alpha$ even down to $\tau_\alpha\simeq
T/10$. Therefore a finer discretization was not deemed necessary.

For the two-state driving protocol (Fig.~\ref{fig:generic}A), the mean ligand
number and total receptor number were set to $\lambda_0=N=25$. The ligand death
rate was set to $\beta=100$, and the mean driving period was set to $T=2\pi$ in
simulation time units. Switching times were generated independently, following a
Gamma- (or Erlang-) distribution with shape parameter $k\in\{1,2,3,5,10,20,50\}$
and mean $T/2$. The input rate was set to a random initial value $\alpha_0[1\pm
\rho]$ and then toggled after each random switching time between $\alpha_0[1\pm
\rho]$, where $\alpha_0=\lambda_0\beta$ and $\rho=0.5$. For a given ligand
dissociation rate constant $\mu$, the association constant was set to
$\gamma=\mu/\lambda_0$ to ensure half-filling at the average driving rate. 

For the harmonic-oscillator protocol (Fig.~\ref{fig:generic}B) the same
parameters were used, except that the driving signal was now generated by a
forward-Euler integration of the Langevin equation given in the main text. The
damping parameter $\eta$ was varied in $\{1/2,1,2,4\}$.

For each value of $\mu$, the system state was initialized to the equilibrium
molecule numbers at $\alpha=\alpha_0$, and $N_\mathrm{tr}=2000$ trajectories of
length $10T$ were generated. The dissipated heat contributions $-\Delta\Phi$
(Eq.~\ref{eq:heat}) were accumulated on every reaction and averaged to yield
$\Qbind$.

Trajectories were sampled at discrete time intervals $T/100$, and the
corresponding samples of the input rate were binned. The input-output mutual
information was estimated by applying the definition Eq.~\ref{eq:defI} to the
binned simulation data. In doing so, the choice of bin size for
the continuous variable $\alpha$ (in the harmonic-oscillator case) can lead to
systematic errors; we found the results for $I$ to be independent of the bin
size in a plateau region around $N_\mathrm{bin}=100$ equally filled bins, and
therefore used this binning for Fig.~\ref{fig:generic}B.

The data were split into 10 blocks of 200 trajectories each and the mutual 
information $I[n,\alpha_\tau]$ for various prediction intervals $\tau\in [0, T/
2]$
were calculated based on histograms of the discrete-valued samples, for each
block. Plots show the averages over blocks together with standard errors of
the mean estimated from block-wise variation.

\end{document}